\documentclass[twocolumn,aps,prb,floatfix,superscriptaddress]{revtex4}
\usepackage{amsmath,amssymb}
\usepackage{graphicx}
\DeclareMathAlphabet{\bi}{OML}{cmm}{b}{it}

\newcommand{\bra}{\langle}
\newcommand{\cl}{\chi_{\text{loc}}}
\newcommand{\et}{1-2r+s} 
\newcommand{\half}{\mbox{$\frac{1}{2}$}}
\newcommand{\ket}{\rangle}
\newcommand{\kk}{\bi{k}}
\newcommand{\mimp}{m_{\text{imp}}}
\newcommand{\qq}{\bi{q}}
\newcommand{\rs}{r_+}
\newcommand{\veps}{\varepsilon}
\newcommand{\w}{\omega}
\newcommand{\ra}{\to}
\newcommand{\be}{\begin{equation}}
\newcommand{\ee}{\end{equation}}
\newcommand{\bea}{\begin{eqnarray}}
\newcommand{\eea}{\end{eqnarray}}

\begin{document}

\title{
Quantum phase transitions in a resonant-level model with dissipation: \\
Renormalization-group studies
}

\author{Chung-Hou Chung}
\affiliation{Electrophysics Department, National Chiao-Tung University,
Hsinchu, Taiwan, R.O.C.}
\author{Matthew T. Glossop}
\affiliation{Department of Physics, University of Florida, Gainesville,
FL 32611-8440, USA}
\author{Lars Fritz}
\affiliation{Institut f\"ur Theoretische Physik,
Universit\"at K\"oln, Z\"ulpicher Stra\ss e 77, 50937 K\"oln, Germany}
\affiliation{Department of Physics, Harvard University, Cambridge MA 02138, USA}
\author{Marijana Kir\'can}
\affiliation{Max-Planck-Institut f\"ur Festk\"orperforschung,
Heisenbergstra\ss e 1, 70569 Stuttgart, Germany}
\author{Kevin Ingersent}
\affiliation{Department of Physics, University of Florida, Gainesville,
FL 32611-8440, USA}
\author{Matthias Vojta}
\affiliation{Institut f\"ur Theoretische Physik,
Universit\"at K\"oln, Z\"ulpicher Stra\ss e 77, 50937 K\"oln, Germany}

\date{\today}

\begin{abstract}
We study a spinless level that hybridizes with a fermionic band and is also
coupled via its charge to a dissipative bosonic bath. We consider the general
case of a power-law hybridization function $\Gamma(\w)\propto |\w|^r$ with
$r\ge 0$, and a bosonic bath spectral function $B(\w)\propto \w^s$ with
$s\ge -1$. For $r<1$ and $\mathrm{max}(0,2r-1)<s<1$, this Bose-Fermi quantum
impurity model features a continuous zero-temperature transition between a
delocalized phase, with tunneling between the impurity level and the band,
and a localized phase, in which dissipation suppresses tunneling in the
low-energy limit. The phase diagram and the critical behavior of the model are
elucidated using perturbative and numerical renormalization-group techniques,
between which there is excellent agreement in the appropriate regimes. For
$r=0$ this model's critical properties coincide with those of the spin-boson
and Ising Bose-Fermi Kondo models, as expected from bosonization.
\end{abstract}


\maketitle


\section{Introduction}
\label{sec:intro}

Quantum phase transitions\cite{subirbook} in mesoscopic systems form a growing
area of condensed matter research. From a theoretical perspective, it is known
that models of a finite system (the ``impurity'') coupled to infinite baths may
exhibit boundary quantum phase transitions (QPTs), at which
only a subset of the degrees of freedom becomes critical.\cite{mvreview}
Such models help to advance our understanding of quantum criticality
in strongly correlated systems:
Concepts and solution techniques developed in the impurity context may
be applied to lattice models, e.g., within the framework of dynamical mean-field
theory (DMFT)\cite{dmft-rmp} and its extensions.
This approach has been followed in connection with the ``local criticality''
proposed to underlie the anomalous non-Fermi-liquid behavior of
several heavy-fermion systems.\cite{lcqpt}
On the experimental side, QPTs in mesoscopic few-level
systems are of great interest, both for the unprecedented opportunity to probe
quantum criticality in a direct and highly controlled fashion,\cite{potok,dias}
and for their numerous potential technological applications, e.g., in
nanoelectronics and quantum information processing.\cite{kondo,device,noise}

In recent years, QPTs have been identified and
studied in a number of quantum impurity models.\cite{mvreview} Such models
can contain both fermionic bands (e.g., conduction-electron quasiparticles)
and bosonic baths (e.g., phonons, spin fluctuations, or electromagnetic noise).
Analytical and numerical techniques have been refined to analyze the critical
behavior of these models. Analytical approaches based on bosonization
or conformal field theory have been used extensively, although their
applicability is limited, e.g., to certain forms of the bath spectrum.
For other situations, powerful epsilon-expansion techniques have been
developed. As such expansions are asymptotic in character, a comparison with
numerical results is mandatory to assess their reliability.

An example with especially rich behavior is the fermionic pseudogap Kondo
model,\cite{withoff} which features QPTs between Kondo-screened and
local-moment ground states.\cite{withoff,cassa,GBI,insi,lars,larslong}
Essentially perfect agreement between the results of various epsilon expansions
(around different critical dimensions) and numerical renormalization-group (NRG)
calculations has been found in critical exponents as well as universal
amplitudes such as the residual impurity entropy.\cite{lars,larslong}

Impurity models that include bosons are harder to tackle numerically than
pure-fermionic problems due to the large Hilbert space, and fewer results are
available.  The development of a bosonic version\cite{BTV,BLTV} of Wilson's
NRG approach\cite{Bulla:07} has made possible a detailed nonperturbative study
of the spin-boson model, where tunneling in a two-state system competes with
dissipation.\cite{leggett} For the case of Ohmic dissipation, the spin-boson
model has long been known to display a QPT of the Kosterlitz-Thouless type.
In the sub-Ohmic case, the model instead exhibits a line of continuous QPTs
governed by interacting quantum critical points (QCPs).\cite{BTV,BLTV,VTB}
(The latter lie in a different universality class than the QCP of the pseudogap
Kondo model.)

Of particular interest, both for mesoscopics and in the context of extended
DMFT for correlated lattice-systems,\cite{edmft,chitra} are impurity models
with fermionic \textit{and} bosonic baths. The best-studied member of this
class is the Bose-Fermi Kondo model,\cite{bfk,sengupta,bfknew,kircan2,kirchner}
with a spin-$\half$ local moment coupled to fermionic quasiparticles
(the regular Kondo model) as well as to a bosonic bath.
The latter may describe spin or charge fluctuations of the bulk system
in which the impurity is embedded.
The scope of NRG applications has recently been widened to provide a
comprehensive treatment of an Ising-symmetric
version of the Bose-Fermi Kondo model.\cite{Glossop:05,Glossop:07}

The purpose of this paper is to investigate a somewhat simpler quantum
impurity model containing both fermionic and bosonic baths, namely a
resonant-level model of spinless electrons, with the impurity charge
coupled to a dissipative reservoir. In standard notation, its Hamiltonian is
\begin{align}
\label{H}
\mathcal{H}& = \veps_f f^\dagger f + \sum_{\kk}v^{\phantom{\dagger}}_{\kk}
  \left( f^\dagger c^{\phantom{\dagger}}_{\kk} + \text{H.c.}\right)
  \,+ \, \sum_{\bf{k}} \veps^{\phantom{\dagger}}_{\kk} c^{\dagger}_{\kk}
  c^{\phantom{\dagger}}_{\kk} \nonumber \\
& \quad + (f^{\dagger} f -\half) \sum_{\qq} g^{\phantom{\dagger}}_{\qq}
  (b^{\phantom{\dagger}}_{\qq} + b_{-\qq}^{\dagger}) + \sum_{\qq}
  \w^{\phantom{\dagger}}_{\qq} b_{\qq}^{\dagger} b^{\phantom{\dagger}}_{\qq}\, ,
\end{align}
with $v_{\kk}$ characterizing the hybridization between
conduction electrons of energy $\veps_{\kk}$ and the impurity level
at energy $\veps_f$, and $g_{\qq}$ coupling bosons of energy
$\w_{\qq}$ to the impurity occupancy. Without loss of generality,
$v_{\kk}$ and $g_{\qq}$ are taken to be real and non-negative.
Equation \eqref{H} represents perhaps the simplest nontrivial Bose-Fermi
quantum impurity model, making it a paradigm for this class and an ideal
problem for detailed comparison between analytical and numerical results.

The model is completely specified by the impurity level energy $\veps_f$,
the hybridization function
\begin{align}
\Gamma(\w)& \equiv \pi\!\sum_{\kk}v_{\kk}^2\delta(\w\!-\!\veps_{\kk})
= \Gamma_0\left|\frac{\w}{D}\right|^r && \!\!\!\!\text{for } |\omega|<D , \\
\intertext{and the bosonic bath spectral function}
B(\w)& \equiv \pi\!\sum_{\qq}g_{\qq}^2\delta(\w\!-\!\w_{\qq})
= B_0\left(\frac{\w}{\w_c}\right)^{\!\!s} && \!\!\!\!\text{for }  0<\w<\w_c ,
\end{align}
with $D$ and $\w_c$ acting as fermionic and bosonic cutoffs, respectively.
Thus, in addition to a power-law spectrum for the bosonic bath density of
states (DOS) characterized by an exponent $s$, we consider a nonconstant particle-hole (p-h) symmetric
hybridization function characterized by an exponent $r$. Increasing $r$ (and hence depleting
the hybridization function around the Fermi level $\w=0$) and increasing $B_0$
both act to suppress tunneling between the local level and the conduction band.
For most of the numerical work presented in Sec.\ \ref{sec:NRG},
we fix $r$, $s$, and the hybridization strength $\Gamma_0$, then tune the
dissipation strength $B_0$ to the vicinity of a QPT.

Although the bath densities of states and $v_{\kk}$, $g_{\qq}$
do not require separate specification, it will facilitate comparison
between numerical and perturbative results to assume that $v_{\kk}=v_0$,
$g_{\qq}=g_0$ for all $\kk$, $\qq$. In this case,
$\Gamma(\w)=\pi v_0^2 \rho_c(\w)$ and $B(\w)=\pi g_0^2 \rho_b(\w)$, with
the fermionic and bosonic DOS given, respectively, by
\begin{align}
\rho_c(\w)& = N_0\left|\w/D\right|^r &&  \text{for } |\omega|<D \, ,
\label{fdos} \\[1ex]
\rho_b(\w)&
= (K_0^2/\pi)\left(\w/\w_c\right)^s && \text{for } 0<\w<\w_c \, ,
\label{bdos}
\end{align}
where $N_0$ and $K_0$ are normalization factors.
Thus, $\Gamma_0=\pi N_0 v_0^2$ and $B_0=(K_0g_0)^2$. The metallic case is
recovered for $r=0$, and Ohmic dissipation corresponds to taking $s=1$.

It is convenient to identify a pseudospin---making clear the close
relationship between model \eqref{H} and the spin-boson model and its
variants---by writing
\be
f^{\dagger} \equiv S^{+} , \quad
f \equiv S^{-} , \quad
f^{\dagger} f -\frac{1}{2} \equiv S_z \, .
\label{pseudospin}
\ee
In the model described by Eq.\ \eqref{H}, the friction caused by the bosonic
bath competes with the resonant tunneling of electrons.  In contrast to the
simpler spin-boson model,\cite{leggett} the tunneling properties are determined
by the hybridization function $\Gamma(\w)$.

\begin{figure}[t]
\centerline{\includegraphics[clip,width=5cm]{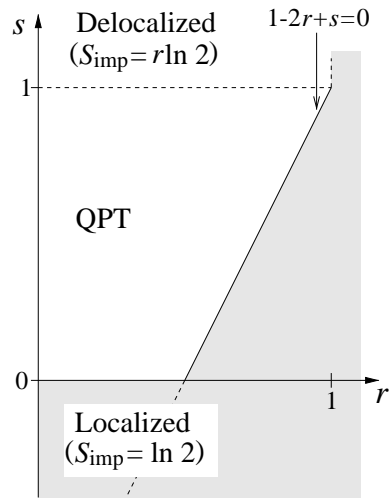}}
\caption{\label{fig:pd}%
Schematic phase diagram of the dissipative resonant-level model
\protect\eqref{H}, in the parameter space spanned by exponents $r$ and $s$
characterizing the low-energy behavior of fermionic and bosonic baths,
respectively. (A finite coupling to both baths is assumed.)
For $\text{max}(0,2r-1)<s\leq1$,
the model shows a boundary quantum phase transition
(as the couplings $v_0$ and $g_0$ are varied)
between a delocalized phase and a localized phase.
(The physics along the line $r=0$ is identical to that of the spin-boson model.)
In contrast, for $s>1$ and $r<1$ the system is generically delocalized,
whereas it is localized in the rest of the parameter regime (shaded).
The impurity entropy $S_{\text{imp}}$ is discussed in the text.
Two perturbative RG expansions are employed: around the free-impurity fixed
point, where the expansion is controlled about $r=s=1$
(Sec.\ \protect\ref{sec:RGvg}) and around the resonant-level fixed point,
where the expansion is controlled in $\et$ (Sec.\ \ref{sec:RGg}).
}
\end{figure}

For $\veps_f=0$ the model features a Z$_2$ symmetry of particle-hole type
[assuming $\rho_c(\omega) = \rho_c(-\omega)$ as noted above], namely
$c_\kk \rightarrow c_\kk^\dagger$, $f \rightarrow -f^\dagger$, and
$S_z \rightarrow -S_z$. Then, we expect that the competition between resonant
tunneling and dissipation yields a QPT between a ``delocalized'' phase
($\bra S_z\ket =0$), in which the principal effect of dissipation is to
renormalize the tunneling amplitude, and a ``localized'' phase
($\bra S_z\ket \ne 0$) with a doubly degenerate ground state,
where the tunneling amplitude renormalizes to zero in the low-energy limit.
We note that for the case of a metallic fermionic bath [$r=0$ in
Eq.\ \eqref{fdos}], bosonization techniques can be used to map the model
\eqref{H} to the spin-boson model.\cite{hur1}
(The same applies to the Ising-symmetric Bose-Fermi Kondo model
with $r=0$, and this equivalence has been verified using
NRG.\cite{Glossop:05,Glossop:07})

In this paper, we employ renormalization-group (RG) techniques to map out the
phase diagram of the Hamiltonian \eqref{H} and to establish over what range of
bath exponents $r$ and $s$ the model can be tuned to a delocalized-to-localized
QPT, akin to that of the spin-boson model. We do so using both perturbative RG
methods, based on epsilon-expansion techniques developed in the context of the
pseudogap Kondo and Anderson models,\cite{larslong} and the Bose-Fermi
extension\cite{Glossop:05,Glossop:07} of the NRG approach, which allows us to
access the entire parameter range of the model.

Our main result is summarized in Fig.\ \ref{fig:pd}, which illustrates the
qualitative behavior of the model in the plane spanned by the bath exponents
$r$ and $s$. A delocalized-to-localized transition---which for $r=0$ is
identically that of the spin-boson model---is present at $r>0$ as well.
A more detailed discussion is given in Sec.\ \ref{sec:RGphase}.

The remainder of the paper is organized as follows.
The perturbative RG analysis is outlined in Sec.\ \ref{sec:RG}, where
results for various critical exponents are obtained by expansion around
two distinct fixed points. In Sec.\ \ref{sec:NRG}, we provide
nonperturbative NRG results for the model, including discussion of the
phase diagram, the response to a local field, and the single-particle
spectral function.
We find excellent quantitative agreement between analytical and numerical
results in the appropriate limits. Although the critical properties of the
model \eqref{H} for $r=0$ are established via the mapping to the
spin-boson model, we confirm the equivalence by direct calculation.


\section{Perturbative renormalization group}
\label{sec:RG}

\subsection{Zero-temperature phases}

We begin by discussing the trivial fixed points of the model \eqref{H}
in the presence of p-h symmetry, $\veps_f=0$.
As a characterization, we will refer to the residual impurity entropy
$S_{\rm imp}$, which is defined as the impurity contribution to
the total entropy in the limit temperature $T\to 0$.\cite{mvreview}

For $v_0=g_0=0$, the impurity is decoupled from both baths.
We denote this free-impurity fixed point by FImp.
The ground state is doubly degenerate: $S_{\text{imp}}=\ln 2$.

For $v_0\neq 0$ and $g_0=0$ one has a resonant-level model with a
power-law conduction-band DOS given by Eq.\ \eqref{fdos}.
The hybridization is relevant in the RG sense (w.r.t. FImp) for $r<1$,
and hence the impurity charge strongly fluctuates.\cite{GBI,larslong}
We refer to this as the delocalized fixed point (Deloc), which, as
discussed in Ref.\ \onlinecite{larslong}, is located at intermediate
RG coupling, $(g,v) = (0,v^\ast)$.
Somewhat surprisingly, the impurity entropy is $S_{\text{imp}}=r \ln 2$,
and vanishes only in the metallic case $r=0$.
For $r>1$, by contrast, the hybridization is RG-irrelevant, and the delocalized
fixed point merges with FImp.\cite{spinfoot}

The dissipative coupling $g_0$ turns out to be RG-relevant at the FImp fixed
point for $s<1$ (see, e.g., Refs.\ \onlinecite{leggett} and \onlinecite{BTV}).
It tends to suppress tunneling in the low-energy limit.
By analogy with the spin-boson model, this can be expected to result
in a doubly degenerate ground state, $S_{\text{imp}}=\ln 2$,
i.e., a phase with broken $Z_2$ symmetry.
This localized fixed point (Loc) corresponds to coupling values
$(g,v) = (\infty,0)$.
(Note that for $s>1$ the effect of the bosonic bath is weak, not
causing localization.)

The preceding discussion suggests that, for $r<1$ and $s<1$, a QPT separates a
delocalized (small-dissipation) phase from a localized (large-dissipation)
phase. Clearly, this applies only to the case of p-h symmetry, $\veps_f=0$.
Otherwise the $Z_2$ symmetry of the Hamiltonian is broken from the outset,
and the phase transition upon variation of the dissipation strength
will be smeared into a crossover; this is analogous to
the behavior of the spin-boson model in the presence of a finite bias.
Furthermore, in situations where the system is localized at $\veps_f=0$,
there will be a first-order transition upon tuning $\veps_f$ from positive
to negative values (as in an ordered magnet subject to a field).

We now proceed with an RG treatment of the model \eqref{H}, carried out
without recourse to bosonization. We can access quantum-critical
properties via two distinct expansions: (i) an expansion around the
free-impurity fixed point (Sec.\ \ref{sec:RGvg}), which is formally valid
provided that the couplings to both baths are small, and (ii) an expansion
around the resonant-level fixed point (Sec.\ \ref{sec:RGg}), performed after
exactly integrating out the $c$ fermions. The second approach proves to
have the wider range of applicability.


\subsection{RG expansion around the free-impurity limit}
\label{sec:RGvg}

In this subsection, we apply an RG epsilon expansion for
weak couplings near the free-impurity fixed point where
$v_0=g_0=0$.

\subsubsection{RG equations}
\label{sec:RGvgeq}

We model the bosonic bath by a relativistic scalar field, $\phi = b+b^\dagger$,
in $d=2+s$ dimensions, with the action
\be
\mathcal{S}_{\phi}= \int_0^{\beta} \!\! d\tau \int^{\Lambda_q} \!\!
\frac{d^{d}\qq}{(2\pi)^{d}} \, \phi_{-\qq} (\tau)
\left (-\partial_\tau^2+ \qq^2 \right) \phi_\qq (\tau) ,
\ee
$\Lambda_q$ being a momentum-space cutoff (related to the energy cutoff
$\omega_c$ via $\omega_c=c \Lambda_q$ with $c=1$ being a velocity).
This produces a DOS of the form
\be
\rho_\phi(\w)=\text{sgn}(\w)  \frac{S_{2+s}}{2}
|\omega|^s =\text{sgn}(\w) \frac{K_0^2}{\pi}
\left|\frac{\w}{\omega_c}\right |^s ,
\ee
for $|\omega|<\omega_c$, with $S_d=2/[(4\pi)^{d/2}\Gamma(d/2)]$.
[Note that $\rho_\phi$ is just a symmetrized version of $\rho_b$ defined in
Eq.\ \eqref{bdos}.]
Similarly, we represent the fermionic bath by Dirac fermions in $(1+r)$
dimensions:
\be
\mathcal{S}_c = \int_0^\beta \!\! d \tau \int_{-\Lambda_k}^{\Lambda_k}
\frac{d k |k|^r}{(2\pi)^{1+r}} \, {\bar c}_{k} (\partial_\tau + k) c_{k} \, ,
\ee
with
$\Lambda_k=D/v_F$ and $v_F=1$ being the (Fermi) velocity,
which reproduces the DOS defined in Eq.\ \eqref{fdos}.
A path-integral representation of Eq.\ \eqref{H}
reads
\begin{align}
\label{rlvact}
\mathcal{S}& = \mathcal{S}_c + \mathcal{S}_\phi
+ \int_0^\beta \!\! d \tau \bar{f} \: \partial_\tau f
+ g_0 \int_0^\beta \!\! d \tau \, ({\bar f} f - \half) \, \phi (\tau,0)
\nonumber \\
& \quad + v_0 \int_0^\beta \!\! d \tau
\left[ \bar{f}c (\tau,0) + \text{c.c.} \right] .
\end{align}
Power counting yields the bare scaling
dimensions of fields and couplings with respect to $v_0=g_0=0$:
$[f]=0$, $[\phi_{\qq}]=-(1+s)/2$, $[c_\kk]=-(1+r)/2$,
$[v_0]=(1-r)/2$, and $[g_0]=(1-s)/2$.
Thus, we can carry out an RG expansion around $r = 1$ and
$s = 1$, where both $v_0$ and $g_0$ become marginal, defining
\be
\epsilon=\frac{1}{2}(1-s), \quad
\epsilon'=\frac{1}{2}(1-r).
\ee

\begin{figure}[t]
\centerline{\includegraphics[clip,width=7.5cm]{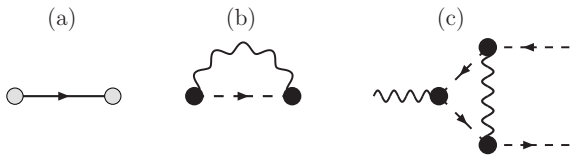}}
\caption{\label{spinless}%
Diagrams appearing in the perturbative expansion for the dissipative
resonant-level model. Dashed, solid, and wiggly lines denote respectively
$f$, $c$, and $\phi$ propagators. The gray (black) circles
are the interaction vertices $v$ ($g$).
(a) and (b): $f$ fermion self-energy diagrams to one-loop order.
(c) One-loop vertex renormalization of $g$.}
\end{figure}

In order to proceed with the RG analysis, we define a renormalized
field $f_R$ and couplings $v$ and $g$ according to
\be
\label{Zfgspinless}
\begin{split}
f & = \sqrt{Z_f} f_R \, , \\
v_0 & = \mu^{\epsilon'} \sqrt{\frac{D^r}{N_0 Z_f}} Z_v v \, , \\
g_0 & = \mu^{\epsilon}\frac{ \sqrt{\omega_c^s \pi} Z_g}{K_0 Z_f} g \, ,
\end{split}
\ee
where $\mu$ is an arbitrary renormalization energy scale and $Z_f$, $Z_v$, and $Z_g$ are
renormalization factors.  As is usual for impurity problems, there is no
renormalization of the bosonic and fermionic bulk propagators, since
the impurity only provides a one-over-volume correction to the bulk
properties. The relevant diagrams for obtaining the one-loop RG beta
functions are shown in Fig.\ \ref{spinless}.

Following standard procedures,\cite{bgz} the one-loop RG beta functions of the
dissipative resonant-level model are given by
\be
\label{betaspinless}
\begin{split}
\beta (v) & = -\epsilon' v+ v^3+\half g^2 v \, , \\
\beta (g) & = - \epsilon g + 2 v^2 g \, ,
\end{split}
\ee
where the calculation parallels that of Ref.\ \onlinecite{larslong}.
The corresponding $Z$ factors, to one-loop accuracy, are
$Z_f=1-v^2/\epsilon'-g^2/2\epsilon$, $Z_v=1$, and
$Z_g=1-g^2/2\epsilon$.

\begin{figure}[t]
\centerline{\includegraphics[clip,width=4.5cm]{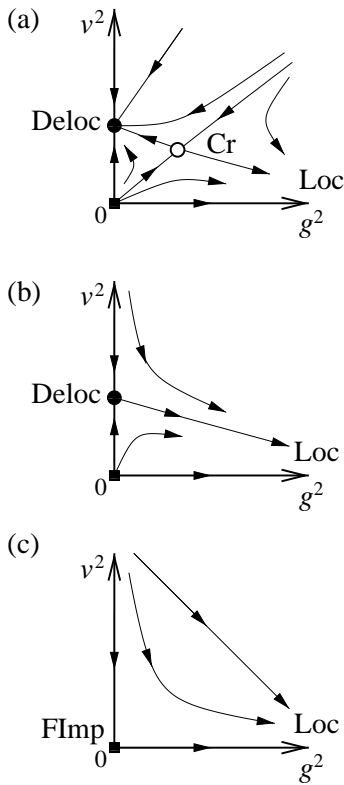}}
\caption{\label{fig:flow_vg}%
Schematic RG flow diagrams for the dissipative resonant-level model with
p-h symmetry. Although these diagrams are obtained by expansion about the
free-impurity fixed point and hence are formally valid as $r,s \to 1$,
they are confirmed by expansion about the delocalized fixed point (Sec.\
\protect\ref{sec:RGg}) and NRG calculations (Sec.\ \protect\ref{sec:NRG})
to capture the correct physics for all $r\ge 0$ and $0<s<1$.
The horizontal axis denotes the renormalized bosonic coupling $g$;
the vertical axis denotes the renormalized hybridization $v$.
(a) $r<\rs=(1+s)/2$: Stable fixed points at Deloc and Loc describe the
delocalized and localized phases, respectively. The continuous impurity
QPT is controlled by the critical fixed point Cr.
(b) $\rs<r<1$:  The delocalized (Deloc) fixed point
is unstable against finite $g$.
As $r\to 1^-$, the Deloc fixed point merges with the free-impurity
fixed point (FImp).  (c) $r\geq 1$: $v$ is irrelevant.
In both (b) and (c), the flow is toward Loc for any finite $g$.}
\end{figure}

The RG flows arising from Eqs.\ \eqref{betaspinless} are plotted
in Fig.\ \ref{fig:flow_vg}. In this subsection, we consider the case
$0<s<1$; the regime $s<0$ is discussed in Sec.\ \ref{sec:RGsneg}.
Fixed points at $(g^{\ast\,2}, v^{\ast\,2})= (0, \epsilon')$
and $(g^{\ast\,2}, v^{\ast\,2}) = (\infty, 0)$ describe the
delocalized (Deloc) and localized (Loc) phases, respectively.
For $r<\rs$, where
\be
\rs = (1+s)/2 ,
\label{rs}
\ee
both these fixed points are stable:
For small $g_0$ and large $v_0$, the ground state is delocalized,
characterized by strong local charge fluctuations due to resonant
tunneling between the impurity and the conduction electron bath
($\langle S_z \rangle = 0$).
In the opposite limit of small $v_0$ and large $g_0$, we find a
localized ground state where charge tunneling renormalizes to zero
in the low-energy limit ($\langle S_z \rangle \neq 0$).
An unstable critical fixed point [Cr], located at
$(g^{\ast\,2}, v^{\ast\,2})= (2\epsilon'-\epsilon,\epsilon/2)$,
controls the QPT between these two phases.  This critical
fixed point lies on the separatrix specifying the phase boundary
in the $g_0$-$v_0$ plane between the delocalized and localized phases.

As $r$ approaches $\rs$ from below, the critical fixed point merges
with the delocalized fixed point (which itself merges with FImp
as $r\to1$ from below).
Hence, no transition occurs for $r\geq\rs$: Deloc and FImp are
unstable w.r.t. infinitesimal bosonic coupling, such that
the ground state is always localized for $g_0 \neq 0$.

\subsubsection{Correlation-length exponent}
\label{sec:RGnu}

In the following, we discuss the properties of the boundary QPT,
controlled by the critical fixed point Cr.
We start with the correlation-length exponent $\nu$, describing the flow away from
criticality: The characteristic energy scale $T^{*}$
above which quantum-critical behavior is observed vanishes as\cite{subirbook}
\be
T^{*} \propto |t|^{\nu} \, ,
\label{nudef}
\ee
where $t$ is a dimensionless measure of the distance to criticality, defined
such that $t>0$ ($t<0$) corresponds to the localized (delocalized) phase.
Upon linearization of the RG beta functions around the Cr fixed point,
we obtain
\be
\label{nuspinless}
\frac{1}{\nu} = \sqrt{\frac{\epsilon^2}{4} +
4\epsilon \left( \epsilon' -\frac{\epsilon}{2} \right)}
-\frac{\epsilon}{2}
+\mathcal{O} \left(\epsilon^2, \epsilon'^{\,2} \right) \, .
\ee
Clearly, $\nu$ diverges as $s\to 1$ and $r\to 1$ together.
By expanding the square-root in Eq.\ \eqref{nuspinless}, the inverse
correlation length exponent can be approximated as $\nu^{-1}=\et$.
The same result, valid for small $\et$, is also obtained in
Sec.\ \ref{sec:RGg} following an RG expansion valid near the strong-coupling
fixed point. The divergence of $\nu$ as $\et\to 0$ is demonstrated
numerically in Sec.\ \ref{sec:NRGnu} and the form compared to
Eq.\ \eqref{nuspinless}.

\subsubsection{Response to a local field}
\label{sec:RGlocresp}

The local impurity susceptibility $\cl(T)$ is the impurity response
to a field applied only to the impurity.\cite{mvreview}
Here, for the spinless resonant-level model under consideration, the level
energy $\epsilon_f$ plays the role of a local electric field.
Defining the impurity ``magnetization'' $\mimp= \langle S_z \rangle$, with the
pseudospin $S_z$ as specified in Eq.\ \eqref{pseudospin}, it follows that
\be
\cl = -\frac{\partial \mimp}{\partial \veps_f}
\ee
is nothing other than the impurity capacitance.

Near criticality, $\cl(T)$ is expected to follow a power-law form
\be
\cl(T) \propto \frac{1}{T^{1-\eta_{\chi}}} \quad
\text{for } T^*\ll T \ll T_0 ,
\label{chiloc}
\ee
up to a nonuniversal cutoff scale $T_0$.
This relation defines the anomalous exponent $\eta_{\chi}$, which governs
the anomalous decay of the impurity ``spin-spin'' correlation function
and is calculated via
\be
\eta_\chi = \mu \frac{\partial \ln Z_\chi}%
{\partial \mu}\bigg|_{v^\ast, g^\ast} \, .
\ee
The renormalization factor $Z_{\chi}$ obeys the exact
relation\cite{mvreview,bfknew}
\be
\label{eq:zchi}
Z_\chi^{-1}= (Z_g/Z_f)^2 \, ,
\ee
which is graphically represented in Fig.\ \ref{fig:exact}(a).
This allows us to derive the exact result
\be
\label{etachi:spinless}
\eta_{\chi} = 2\epsilon = 1-s
\ee
at the Cr fixed point, a relation that is borne out by the numerical
results presented in Sec.\ \ref{sec:NRG}.

\subsubsection{Conduction electron $T$-matrix}
\label{sec:RGtmatrix}

\begin{figure}[t]
\centerline{\includegraphics[clip,width=7.5cm]{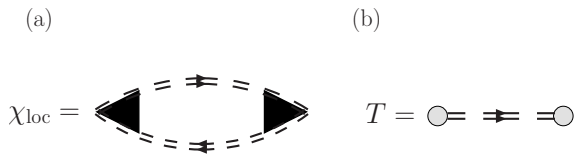}}
\caption{\label{fig:exact}%
(a) Exact relation for the local susceptibility. The black triangle
denotes the full vertex function and the dashed double line denotes
the full impurity level propagator.
(b) The large dot denotes the full hybridization vertex.}
\end{figure}

The conduction electron $T$-matrix, describing the scattering of the $c$
electrons off the impurity, is another important observable, being central
to the calculation of transport properties. For a resonant-level model, the
$T$-matrix is given by $T(\omega) = v_0^2 G_f(\omega)$ where $G_f$ is the full
impurity ($f$-electron) Green's function, graphically represented in
Fig.\ \ref{fig:exact}(b). As with the local susceptibility, we expect a
power-law behavior of the $T$-matrix spectral density near criticality:
\be
T(\omega) \propto \frac{1}{|\omega|^{1-\eta_T}} \quad
\text{for } T^*\ll |\omega|\ll T_0 \, .
\label{tmatrix}
\ee
It has been shown\cite{larslong} that all critical fixed points for
$0<r<1$ in the pseudogap Anderson and Kondo models display
$T(\omega) \propto |\omega|^{-r}$ as $\w \ra 0$, which behavior has been
observed in a number of separate
studies.\cite{Bulla:97,Logan:00,Bulla:00}

Using the exact relation $Z_T = Z_f/Z_v^2$, we can derive an exact result
for the critical point of the dissipative resonant level model:
\be
\eta_{T}= 1-r \, .
\label{etaTweak}
\ee
Thus, even though the multiplicative prefactor of the behavior \eqref{tmatrix}
is expected to exhibit both $r$ and $s$ dependence, the power law followed at
criticality is identical to that of the pseudogap Kondo and Anderson models.

\subsubsection{Hyperscaling and other critical exponents}
\label{sec:RGhyper}

The QCP is expected to satisfy hyperscaling relations characteristic of an
interacting fixed point, including $\w / T$ scaling in dynamical
quantities.\cite{mvreview}
It follows that the correlation-length exponent $\nu$ and
the anomalous exponent $\eta_{\chi}$ are sufficient to determine all critical
exponents associated with the application of a local field.\cite{insi,mvreview}
For example, one can define exponents $\gamma$ and $\gamma'$ through the
$T \to 0$ limit of the local susceptibility near criticality:
\be
\begin{gathered}
\cl (t<0; T=0) \propto (-t)^{-\gamma}, \;\;\;
\gamma =\nu(1-\eta_{\chi}) \, , \\
T\cl (t>0; T=0) \propto t^{\gamma^{\prime}}, \;\;\;
\gamma^{\prime} =\nu\eta_{\chi} \, .
\end{gathered}
\ee
One can also determine critical exponents $\beta$ and $\delta$
associated with the local magnetization $\mimp$:
\be
\label{betadelta}
\begin{gathered}
\mimp(t>0; T=0,\veps_f\to 0) \propto t^{\beta}, \;\;\;
\beta = \nu \eta_{\chi}/2 , \\
\mimp(\veps_f; t=0, T=0) \propto |\veps_f|^{1/\delta}, \;\;\;
\delta = 2/\eta_{\chi}-1 .
\end{gathered}
\ee
Thus, near criticality
\be
\beta = \frac{\epsilon}{\sqrt{\epsilon^2/4 + 4\epsilon
(\epsilon' -\epsilon/2)} - \epsilon/2}
+\mathcal{O} \left( \epsilon^2, \epsilon'^{\,2} \right)
\label{beta_weak}
\ee
and
\be
\delta = \frac{1}{\epsilon} - 1
+ \mathcal{O} \left(\epsilon^2, \epsilon'^{\,2} \right) ,
\label{delta_weak}
\ee
where, in contrast to Eqs.\ \eqref{etachi:spinless} and \eqref{etaTweak},
the higher-order corrections do not cancel.
Section \ref{sec:NRG} reports NRG results for several of these critical
exponents that demonstrably obey the hyperscaling relations.

\subsection{RG expansion around the delocalized fixed point}
\label{sec:RGg}

In addition to the RG expansion for $r\to 1$ and $s\to 1$, as
described in Sec.\ \ref{sec:RGvg}, a second epsilon expansion can
be performed around the Deloc fixed point.

\subsubsection{RG equations}
\label{sec:RGgeq}

\begin{figure}[t]
\centerline{\includegraphics[clip,width=6cm]{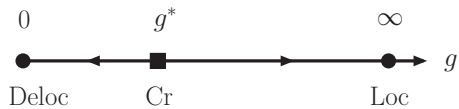}}
\caption{\label{RGflow_strong}%
RG flow diagram of the dissipative resonant-level model near the delocalized
(Deloc) fixed point for $s$, $\tilde{\epsilon}>0$. The two stable phases
are governed by the delocalized ($g=0$) and localized ($g=\infty$)
fixed points, separated by the critical fixed point [$g=g^{\ast}$
specified in Eq.\ \protect\eqref{eq:fixedpoint}].}
\end{figure}

To begin, we integrate out the conduction electrons, which is an exact
operation for the present model. The resulting action is\cite{larslong}
\begin{align}
\label{symact}
\mathcal{S} & =  \sum_{\omega_n} {\bar f}(\omega_n)\, \bigl[ i A_0
\text{sgn}(\omega_n)|\omega_n/D|^r + iA_1\omega_n \bigr] \, f(\omega_n) \nonumber \\
& \quad + \mathcal{S}_\phi + g_0 \int_0^{\beta} \!\! d\tau
\left( {\bar f} f - \half \right ) \phi(\tau,0) \, ,
\end{align}
where the local $f$ fermions are now ``dressed'' by the conduction lines,
\be
A_0=\pi N_0 v_0^2 \sec\left(\frac{\pi r}{2}\right) = \Gamma_0 \sec\left(\frac{\pi
r}{2}\right)
\ee
is a nonuniversal energy scale, and $A_1 = 1+\mathcal{O}(v_0^2)$.
For $r<1$, the $|\omega_n|^r$ term dominates the $f$ propagator
at low energies.
Then, dimensional analysis of the bosonic coupling (here w.r.t. the Deloc fixed point)
yields
\be
[g_0]=\frac{2 r-1-s}{2} \, ,
\ee
which implies that an RG expansion can be controlled in the smallness of
\be
2\tilde{\epsilon} = \et.
\ee
We introduce a dimensionless coupling according
\be
g_0=\mu^{-\tilde{\epsilon}}A_0 \frac{\sqrt{\omega_c^s\pi}Z_g}{K_0 Z_f}g \, ,
\ee
and, following the procedure described in Sec.\ \ref{sec:RGvg}, we find that the
only contribution to $Z_g$ is that shown in Fig.\ \ref{spinless}(c), which reads
(note that $Z_f=1$ to this order)
\be
Z_g=1+\csc\!\left(\frac{\pi s}{2} \right)\frac{g^2}{\tilde{\epsilon}} \, .
\ee
The RG beta function for $g$ is
\be
\label{betastrong}
\beta (g) = \tilde{\epsilon}g - 2 \csc\!\left(\frac{\pi s}{2}\right) g^3 .
\ee
It is clear from Eq.\ \eqref{betastrong} that for $s>0$ and
$\tilde{\epsilon}>0$, there exists a critical fixed point at
\be
\label{eq:fixedpoint}
g^{\ast\,2} = \frac{\tilde{\epsilon}}{2} \sin\left(\frac{\pi s}{2}\right) \, ,
\ee
which controls the delocalized-to-localized transition.
The RG flow diagram is sketched in Fig.\ \ref{RGflow_strong}.

Note that the critical coupling $g^{\ast}$ approaches zero as
$\tilde{\epsilon}\to 0^+$ and/or as $s\to 0^+$, suggesting that
beyond these limiting cases the delocalized fixed point is
unstable towards the localized fixed point.
The same instability has already been deduced for $\tilde{\epsilon}<0$
[i.e., for $r>\rs=(1+s)/2$], based on expansion about the free-impurity
fixed point (see Sec.\ \ref{sec:RGvg}).
The behavior for $s\le 0$ is analyzed in the next section.

\subsubsection{The regime $s \le 0$}
\label{sec:RGsneg}

For $s\le 0$, the perturbation theory described in Sec.\ \ref{sec:RGg} is
singular due to the divergent DOS in the bosonic propagator.
In this range of $s$, the delocalized fixed point is always unstable against
any infinitesimal bosonic coupling $g_0$, which favors the localized fixed
point.

We can gain a better understanding of this instability by
considering the local bosonic propagator $G_{\phi_0}(i\omega_n) =
\sum_\qq G_{\phi}(\qq,i\omega_n)$ in the presence of the impurity.
Including impurity effects via the boson self-energy, the local boson
propagator is given by
\be
G^{-1}_{\phi_0}(i\omega_n) =
\left\{ \begin{array}{ll}
\omega_n^s + s \Lambda^s - g_0^2 \quad & \text{for } s>0, \\[1ex]
\omega_n^{-s} - g_0^2 & \text{for } s\le 0,
\end{array} \right.
\label{Gphi0}
\ee
where $\Lambda$ is a momentum cutoff energy scale.
Let us discuss $s>0$ first.
For $s\Lambda^s > g_0^2 > 0$, the local boson propagator is massive,
meaning that the ground state for the bulk is just the empty state.
For $g_0^2>s\Lambda^s > 0$, by contrast, the local boson propagator has
``negative mass'', as a consequence of which the local boson condenses at zero
temperature with an expectation value $\langle \phi_0 \rangle \neq 0$.
This drives the system to the localized phase where the pseudospin operator
$S_z$ also assumes a nonzero expectation value.
This reasoning supports the existence of a QPT for $s>0$,
with criticality reached at $g^{\ast\,2} = s\Lambda^s$.
For $s\le 0$, the local boson propagator $G_{\phi_0}$ always has a
negative mass, i.e., the impurity is localized.
(Technically, the impurity induces a bound state in $G_{\phi_0}$.)
The observation that the ground state is always localized for $s\le 0$
is consistent with previous studies of
the spin-boson model\cite{BTV,VTB} and the Bose-Fermi Kondo
model,\cite{Glossop:05,Glossop:07} which belong to the same universality
class as the dissipative resonant-level model in the metallic limit $r=0$.

\subsubsection{Phase diagram}
\label{sec:RGphase}

The RG flow allows us to deduce that the qualitative phase diagram of the
dissipative resonant-level model in the parameter space specified by $r$
and $s$ is as shown in Fig.\ \ref{fig:pd}. The solid line denotes the locus
of points satisfying $\et=0$. In the unshaded region to the left of
the line [i.e., for $\text{max}(0,2r-1)<s<1$, or equivalently $\half<\rs<r<1$
with $\rs$ defined in Eq.\ \eqref{rs}], the RG expansion predicts a continuous
QPT as $v_0$ and $g_0$ are varied. For $s<\text{max}(0,2r-1)$ (shaded area),
the ground state of the model is always localized for any finite bosonic
coupling $g_0$. This is consistent with the RG flow diagrams presented in
Fig.\ \ref{fig:flow_vg}, where the RG expansion is carried out for $r,s\to 1$.
The phase diagram is confirmed by NRG results in Sec.\ \ref{sec:NRG}.

\subsubsection{Critical exponents}
\label{sec:RGexponents}

By linearizing the RG equation around the fixed point, the correlation-length
exponent at the critical point $g^{\ast}$ is found to satisfy
\be
\frac{1}{\nu} = 2\tilde{\epsilon}+\mathcal{O} \left(\tilde{\epsilon}^2\right) .
\label{nuspinless2}
\ee

For the anomalous exponent $\eta_{\chi}$ associated with the local
susceptibility [Eq.\ \eqref{chiloc}], we again have the exact property
Eq.\ \eqref{eq:zchi} [see also Fig.\ \ref{fig:exact}(a)], from which it
follows that
\be
\label{etachi:spinless2}
\eta_{\chi} = 1-s \, .
\ee
The exponents $\beta$ and $\delta$ can be obtained from the hyperscaling
relations \eqref{betadelta}:
\be
\beta = \frac{1-s}{4\tilde{\epsilon}}
+ \mathcal{O}\left(\tilde{\epsilon}^2\right) , \label{beta_strong}
\ee
and
\be
\delta = \frac{1+s}{1-s} + \mathcal{O}\left(\tilde{\epsilon}^2\right) .
\label{delta_strong}
\ee
The exponent $\eta_T$, associated with conduction-electron $T$-matrix, is
also found to obey $\eta_T = 1-r$ [see Eq.\ \eqref{etaTweak}].
Of course, all critical exponents for the two RG expansions (one for
$r, s\to 1$ and one for $\et\to 0$) are expected to be compatible since
the expansions describe the same QPT. In the limit $r, s\to 1$, the square
root of Eq.\ \eqref{nuspinless} may be expanded to yield
Eq.\ \eqref{nuspinless2}. The equivalences of Eqs. \eqref{beta_weak} and
\eqref{beta_strong} for $\beta$ and of Eqs.\ \eqref{delta_weak} and
\eqref{delta_strong} for $\delta$ are also readily verified.


\section{Numerical renormalization group}
\label{sec:NRG}

The NRG method\cite{Bulla:07}
has recently been extended to provide nonperturbative results
for the Bose-Fermi Kondo model.\cite{Glossop:05,Glossop:07} In the following,
we implement the same approach for the spinless resonant-level model \eqref{H},
which also involves both fermionic and bosonic baths.

\begin{figure}[t]
\centerline{\includegraphics[clip,angle=270,width=7.5cm]{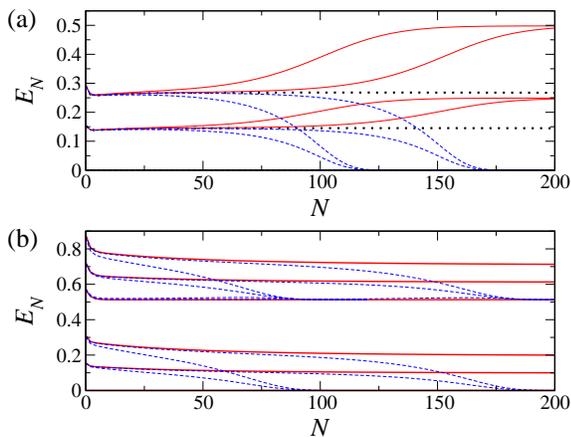}}
\caption{\label{NRGflows}%
(Color online)
(a) The lowest NRG eigenstates $E_N$ vs even iteration number $N$ for
$(r,s)=(0.85,0.9)$, hybridization strength $\Gamma_0=0.1$, and a range of
dissipation strengths $B_0-B_{0,c}=0, \pm 10^{-3}, \pm 10^{-2}$. The flows
are typical of those for $\text{max}(0,2r-1)<s<1$ with $0\le r<1$. The levels
at the critical coupling $B_0=B_{0,c}\approx 0.3731$ are shown as bold dotted
lines while those nearby in the delocalized ($B_0<B_{0,c}$)
[localized ($B_0>B_{0,c}$)] phase are shown as solid [dashed] lines. As $B_0$
approaches $B_{0,c}$ in either phase, the levels follow those of the unstable
critical fixed point down to progressively lower temperatures, before crossing
over to the levels characteristic of the delocalized or localized stable fixed
point. (b) NRG level flows for $(r,s)=(0.975,0.9)$.  In this case, and more
generally for $s<\text{max}(0,2r-1)$ with $0\le r < 1$, the flow is towards
the localized fixed point for any $B_0>0$, but follows the delocalized fixed
point down to progressively lower temperatures as $B_0$ is reduced towards
zero. The solid lines show the flow for $B_0=0$.}
\end{figure}

There are three essential features of the NRG: (i) The energy axis is
logarithmically discretized, introducing a discretization parameter $\Lambda$.
(ii) The Hamiltonian is then mapped to a chain form, with the impurity degrees
of freedom coupled to the first site only of one or more tight-binding chains.
(iii) Owing to the discretization, the tight-binding coefficients decay
exponentially with increasing chain length.  This allows the problem to be
solved in an iterative fashion, diagonalizing progressively longer
finite-length chains and thereby including exponentially smaller energy scales,
$T_N\approx D\Lambda^{-N/2}$, at each iterative step $N=0$, $1$, $2$, $\ldots$.
The RG transformation relating the effective Hamiltonians at consecutive
iterations eventually reaches a scale-invariant fixed point that determines
the low-temperature properties of the system.

In all applications of the NRG, the maximum number $N_s$ of many-body
eigenstates retained from iteration $N$ to form basis states for iteration
$N+1$ must be truncated for sufficiently large $N$ due to the limitations of
finite computational power. The presence of one or more bosonic chains
introduces additional considerations. First, the bosonic Hilbert space must be
truncated even at iteration $N=0$, allowing a maximum of $N_b$ bosons per site
of a bosonic chain. Second, for problems involving both fermionic and bosonic
chains, the fact that the bosonic tight-binding coefficients decay as the
square of those for fermionic chains must be reflected in the specific
iterative scheme employed. That is, only (bosonic and fermionic) excitations
of the same energy scale should be considered at the same iterative step. Thus,
while the fermionic chain is extended at each iteration, the bosonic chain is
extended only at every second iteration. These issues, together with further
details of the implementation of the Bose-Fermi NRG, are discussed in detail in
Ref.\ \onlinecite{Glossop:07}.

The NRG method has provided a comprehensive numerical account of the
quantum-critical properties of a number of impurity problems, e.g., the
fermionic pseudogap Kondo and Anderson models, the spin-boson model, and the
Bose-Fermi Kondo model.  In all cases it is found that the critical properties
(such as exponents) are insensitive to the discretization parameter $\Lambda$
and converge rapidly with the number of retained states $N_s$. For models
involving bosonic baths, critical exponents also rapidly converge with
increasing bosonic truncation parameter $N_b$. In the following we take
$\Lambda=3$, with all data suitably converged for the choice $N_s=500$ and
$N_b=8$. For convenience we set $D=\w_0=1$.

\subsection{Phase diagram}
\label{sec:NRGphase}

Figure \ref{NRGflows} shows the flow of the lowest NRG eigenstates $E_N$ of
the effective Hamiltonian $H_N$ at even iteration numbers $N$ for two
representative cases for $s>0$: (a) $\et > 0$ and (b) $\et<0$.  Figure
\ref{NRGflows}(a) shows data obtained for $(r,s)=(0.85,0.9)$ and $\Gamma_0=0.1$.
Here, and for any $\et>0$, the flow is schematized by Fig. \ref{fig:flow_vg}(a),
which follows from the perturbative analysis. For $B_0<B_{0,c}$, the NRG flow
is towards the delocalized fixed point, where the spectrum coincides with that
for coupling $B_0=0$ to the bosonic bath. For $B_0>B_{0,c}$ the NRG flow
is towards the localized fixed point, where the spectrum coincides with that for
coupling $\Gamma_0=0$ to the fermionic band.  For $B_0$ close to
$B_{0,c}$, as considered in Fig.\ \ref{NRGflows}(a), the flow in either case is
first towards the critical spectrum. The departure from the critical flow, at a
crossover scale $T^*$ that vanishes at $B_0=B_{0,c}$, is governed by the
correlation-length exponent discussed in Sec.\ \ref{sec:NRGnu}.

Figure \ref{NRGflows}(b) shows NRG level flows for $(r,s)=(0.975,0.9)$ and
$\Gamma_0=0.1$. These flows are typical of those for any $\et<0$ and
correspond to the perturbative RG flows of Fig.\ \ref{fig:flow_vg}(b).
The localized ground state obtains for any $B_0>0$. As $B_0$ is reduced
towards zero, the levels follow those of the delocalized fixed point
(obtained for $B_0=0$) down to progressively lower energy scales.

\begin{figure}[t]
\centerline{\includegraphics[clip,width=6.5cm]{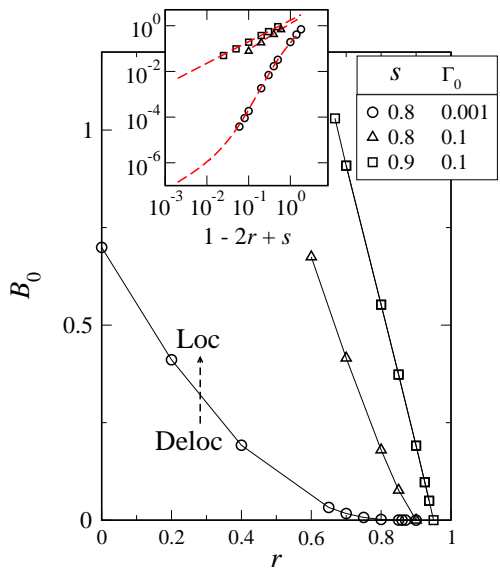}}
\caption{\label{NRG:phdia}%
(Color online)
Phase diagram in the $r$-$B_0$ plane, obtained using NRG for the fixed
bosonic bath exponent $s$ and the hybridization strength $\Gamma_0$
shown in the legend.  For $0<r<\rs=(1+s)/2$, we
find a continuous QPT between delocalized (Deloc) and localized
(Loc) phases. The critical dissipation strength $B_{0,c}$ is
found to vanish continuously at $r=\rs$.  For $r\ge \rs$ only the
localized phase can be accessed for $B_0>0$. The inset shows the
vanishing of $B_{0,c}$ with decreasing $\et$ in each case,
compared to the results obtained from the perturbative analysis.}
\end{figure}

Figure \ref{NRG:phdia} shows the phase diagram of the model on the $r$-$B_0$
plane for three different combinations of the bosonic bath exponent $0<s<1$ and
the hybridization strength $\Gamma_0$. For all $s$ and $\Gamma_0$ pairs
considered, the phase-boundary value of $B_0$ decreases monotonically with
increasing $r$ from that found for a metallic conduction band ($r=0$). This is
particularly clear from the data set obtained for $s=0.8$ and $\Gamma_0=10^{-3}$
(circles in Fig.\ \ref{NRG:phdia}), where the metallic system undergoes a
continuous QPT at a critical $B_{0,c}(r=0)\approx 0.699$. With increasing $r$,
and hence growing depletion of the conduction electron density of states around
the Fermi level, the critical dissipation strength $B_{0,c}$ required to
localize the system is reduced, as expected on physical grounds. $B_{0,c}(r)$
is found to vanish continuously at $r=\rs$, with $\rs$ as defined in
Eq.\ \eqref{rs}. This vanishing is illustrated in the inset to
Fig.\ \ref{NRG:phdia}, which shows $B_{0,c}$ vs $\et$ on a logarithmic scale.

For $r>\rs$, localized solutions are found for arbitrarily small dissipation
strength $B_0>0$. The symbols at the largest $r$ ($=\rs$) in each case, which
lie at $B_0=0$, mark the point at and above which no delocalized solutions
can be found with $B_0>0$. Thus, we find that we can tune the system to a
QPT if, and only if, $0 <s \le 1$ and $0 \le r \le \rs$, in complete agreement
with the scenario deduced via the perturbative analyses and illustrated in
Fig.\ \ref{fig:pd}.

For $0\le r<1$ and $s=1$, we find a line of Kosterlitz-Thouless-like
transitions between delocalized and localized ground states, and for $s>1$
only the delocalized phase is accessed (provided $\Gamma_0>0$). For $r>1$
and $s>1$, the essential physics is controlled by the free-impurity fixed
point, regardless of the couplings $\Gamma_0$ and $B_0$.

\begin{figure}[t]
\centerline{\includegraphics[clip,angle=270,width=7.5cm]{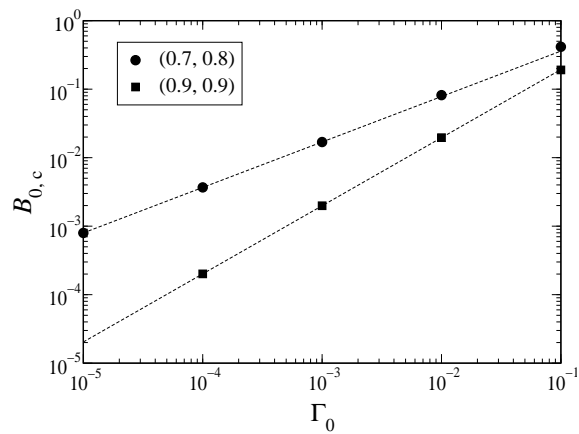}}
\caption{\label{B0cvsdelta}%
Critical dissipation strength $B_{0,c}$ vs hybridization strength
$\Gamma_0$ for the $(r,s)$ pairs specified in the legend. We find
that $B_{0,c}\propto \Gamma_0^x$, with $x=(1-s)/(1-r)$.}
\end{figure}

For a given $(r,s)$ pair that exhibits a continuous QPT, the critical
dissipation strength $B_{0,c}$ varies with the hybridization strength
$\Gamma_0$ as
\be
\label{boundary}
B_{0,c}\propto \Gamma_0^{(1-s)/(1-r)}
\ee
provided that all scales are small compared to the cutoffs. This result, which
follows from dimensional arguments [Eq.\ \eqref{boundary} can readily be
obtained using Eq. \eqref{betaspinless}] and is confirmed numerically in
Fig.\ \ref{B0cvsdelta}, identifies $\Gamma_0^{1/(1-r)}$ as the tunneling
amplitude analogous to $\Delta$ of the spin-boson model, where\cite{BTV} the
critical dissipation strength is $\alpha_c\propto \Delta^{1-s}$. A similar
result for the Bose-Fermi Kondo model finds $B_{0,c}\propto T_K^{1-s}$, with
$T_K$ the bare Kondo temperature serving as a tunneling amplitude between
impurity spin states.\cite{Glossop:05,Glossop:07}

It is interesting to compare the location of the phase boundary obtained using
NRG with that inferred from analytical expansion. We have in mind fixing the
hybridization strength $\Gamma_0$ and the bosonic-bath exponent $s$ (as in
Fig.\ \ref{NRG:phdia}), and finding the critical coupling $B_0$ as a function
of the conduction-band exponent $r$.
However, an analysis of the expansion around the free-impurity fixed point
(Sec.\ \ref{sec:RGvg})
reveals no simple analytical expression for the phase boundary,
due to the fact that the problem is described by a two-parameter flow,
which cannot be linearized in general.
We have therefore analyzed the coupled differential flow equations numerically.
The phase boundary can be obtained by determining the eigenvalues and
eigenvectors of the linearized RG equations near the critical point and then
following the RG flow backwards along the separatrix.

The inset of Fig.\ \ref{NRG:phdia} compares phase boundaries determined via NRG
(symbols) with those obtained via the perturbative RG equations
\eqref{betaspinless} (dashed lines). For the range of $\et$ considered by NRG,
$B_{0,c}$ appears to vanish as a power law, with an exponent that depends on
both the bosonic bath exponent $s$ and the hybridization $\Gamma_0$. This
apparent power law does not reflect the asymptotic behavior, revealed by
the perturbative calculations to be $B_{0,c} \propto \tilde{\epsilon}$ as
$\tilde{\epsilon} \ra 0$. (This regime is inaccessible to NRG because the
merging of the critical and delocalized fixed points with decreasing
$\tilde{\epsilon}$ make it impossible to reliably determine the critical
coupling $B_{0,c}$.)
Nevertheless, we find the level of agreement remarkable and stress that there
is no fitting procedure involved in making this comparison.

From the expansion around the delocalized fixed point (Sec.\ \ref{sec:RGg}),
where we have a one-parameter flow, it seems possible to obtain an analytical
expression for the phase boundary. However, we have to keep in mind that the
dressed $f$ propagator in Eq.\ \eqref{symact} contains terms with different
frequency dependencies, and is dominated by $|\omega_n|^r$ in the low-energy
limit only. (The coefficient $A_1$ is nonzero in general, except right at the
Deloc fixed point.) The interplay of the $|\omega_n|^r$ and $\omega_n$ terms
introduces a nonuniversal crossover scale into the problem, and a proper
treatment including elevated energies would require a multistage RG scheme,
which is beyond the scope of this paper.


\subsection{Critical exponents}

\subsubsection{Correlation-length exponent}
\label{sec:NRGnu}

\begin{figure}[t]
\centerline{\includegraphics[clip,angle=270,width=7cm]{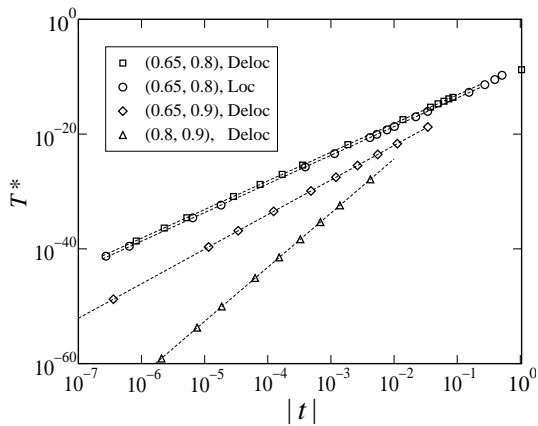}}
\caption{\label{tstar}%
Crossover scale $T^*$ vs $|t|=|B_0-B_{0,c}|/B_{0,c}$ for the $(r,s)$ pairs specified in the legend.
 In the vicinity of the transition ($|t|\ll 1$), $T^*\propto |t|^{\nu}$.  The correlation-length
exponent $\nu(r,s)$ is independent both of the hybridization $\Gamma_0$ and of the phase from which the
QCP is approached.} \end{figure}

The correlation-length exponent $\nu$ defined in Eq.\ \eqref{nudef} is readily
extracted from the crossover scale $T^*\propto \Lambda^{-N^*/2}$ in the NRG
level flows between the unstable and either of the stable fixed points. Here,
$N^*$ denotes the NRG iteration number at which crossover is observed in a
chosen NRG eigenvalue $E_N$. (See Refs.\ \onlinecite{Glossop:05} and
\onlinecite{Glossop:07} for further details.) Figure \ref{tstar} shows $T^*$ vs
$|t|=|B_0-B_{0,c}|/B_{0,c}$ for the $(r,s)$ pairs specfied in the legend. 
The dashed lines are linear fits to the log-log data, which yield the correlation length exponent
$\nu(r,s)$, independent of the hybridization strength $\Gamma_0$ and the phase (Deloc or Loc) from which
the QCP is accessed.

\begin{figure}
\centerline{\includegraphics[clip,angle=270,width=7.5cm]{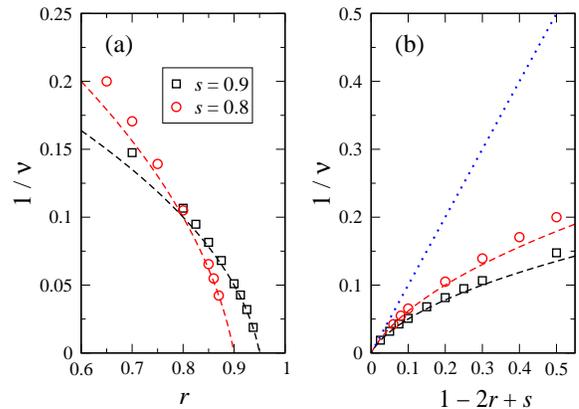}}
\caption{\label{NRG:nu}%
(Color online)
(a) Correlation-length exponent $\nu$ vs conduction-band exponent $r$ for
two values of the bosonic-bath exponent $s$, as shown in the legend. The
symbols show NRG data, while the dashed lines are the corresponding
perturbative results [Eq.\ \protect\eqref{nuspinless}], expanding about the
free-impurity fixed point. We find that $\nu^{-1}$ vanishes at $r=\rs$, in
keeping with the qualitatively distinct behavior for
$2\tilde{\epsilon}\equiv\et\gtrless 0$. (b) The same data plotted vs
$2\tilde{\epsilon}$. For small $\tilde{\epsilon}$,
$\nu^{-1}\approx 2\tilde{\epsilon}$, a result [Eq. \eqref{nuspinless2}]
(shown as a dotted line) obtained by a perturbative expansion
about the delocalized fixed point.}
\end{figure}

The $r$ dependence of the correlation-length exponent is demonstrated in
Fig.\ \ref{NRG:nu}(a) for two values of the bosonic bath exponent $s$.
As anticipated, for $r=0$ we find that within our estimated numerical
error of about 1\%, $\nu(0,s)$ is in essentially exact agreement with
$\nu(s)$ for the spin-boson model\cite{BTV,VTB} (and the Ising-symmetry
Bose-Fermi Kondo model, demonstrated in Ref.\ \onlinecite{Glossop:05} to
share the same universality class).  By increasing $r$ we find that
$\nu(r,s)$ diverges as $r \rightarrow \rs$ from below, i.e., as $\et\to 0^+$.
The dashed lines are the corresponding perturbative results [Eq.\
\eqref{nuspinless}], with which there is excellent agreement for $r$
approaching $\rs$. Figure \ref{NRG:nu}(b) shows the same data plotted
vs $2\tilde{\epsilon}=\et$. With decreasing $\tilde{\epsilon}>0$, the
curves approach the result $\nu^{-1}\approx 2\tilde{\epsilon}$ (shown as a
dotted line), as obtained in Sec.\ II.C.3 by an expansion about the
delocalized fixed point.

\subsubsection{Response to a local field}
\label{sec:NRGlocresp}

\begin{figure}
\centerline{\includegraphics[clip,angle=270,width=7.5cm]{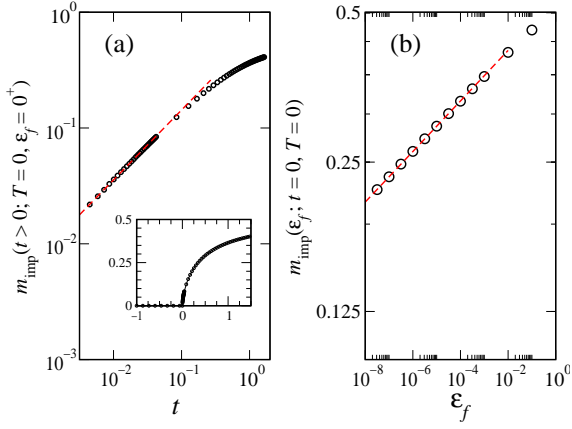}}
\caption{\label{NRG:deltabeta}%
(Color online)
Critical exponents $\beta$ and $\delta$, defined in
Eq.\ \protect\eqref{betadelta}, for $(r,s)=(0.85,0.9)$ and $\Gamma_0=0.1$,
where $B_{0,c}\approx 0.3731$.
(a) Continuous vanishing of order parameter $\mimp$ vs $t=(B_0-B_{0,c})/B_{0,c}$
as $t\to 0^+$ with characteristic exponent $\beta$ (extracted as the limiting
slope of the data on a logarithmic scale). The inset shows the data on an
absolute scale. (b) Variation of $\mimp(T=0)$ with local level energy
$\veps_f$ at the critical point $t=0$. The data clearly follow a power
law for small $\veps_f$, defining the exponent $\delta$.}
\end{figure}

As discussed in Sec.\ \ref{sec:RGlocresp}, the response to a field applied only
at the impurity provides a useful probe of the locally critical properties of
the model.  The inset to Fig.\ \ref{NRG:deltabeta}(a) shows $\mimp(t;T=0)$ vs
$t=(B_0-B_{0,c})/B_{0,c}$ for $(r,s)=(0.85,0.9)$ and hybridization strength
$\Gamma=0.1$. Behaving as a suitable order parameter for the problem,
$\mimp(t; T=0)$ is finite in the localized phase ($t>0$), saturating to
$\mimp(t;T=0)\approx \frac{1}{2}$ for $t\gg 1$ and vanishing continuously as
$t\rightarrow 0^+$. In the delocalized phase ($t<0$), $\mimp(t;T=0)=0$. The
main part of Fig.\ \ref{NRG:deltabeta} shows $\mimp(t;T=0)$ vs $t>0$ on a
logarithmic scale, from which the power-law behavior Eq.\ \eqref{betadelta}
is clearly apparent. The exponent $\beta$ is found to be $\beta=0.601(2)$. At
the QCP ($t=0$), the dependence of $\mimp (t=0,T=0)$ on the field $\veps_f$
defines the exponent $\delta$ according to Eq.\ \eqref{betadelta}.
We typically observe such power-law behavior over several orders of magnitude
of $\veps_f$, as shown in Fig.\ \ref{NRG:deltabeta}(b). For $(r,s)=(0.85,0.9)$,
$1/\delta=0.052(1)$.

We note that for $0\le r <1$ and $s=1$, $\mimp(t;T=0, \veps_f= 0^+)$
undergoes a jump at the critical point $t=0$. Here, the essential behavior
has been discussed in Refs.\ \onlinecite{hur1,Glossop:05,Glossop:07} and
\onlinecite{Borda:05} for the case $(r,s)=(0,1)$ relevant to charge
fluctuations on a metallic island subject to electromagnetic noise.

\begin{figure}
\centerline{\includegraphics[clip,angle=270,width=7cm]{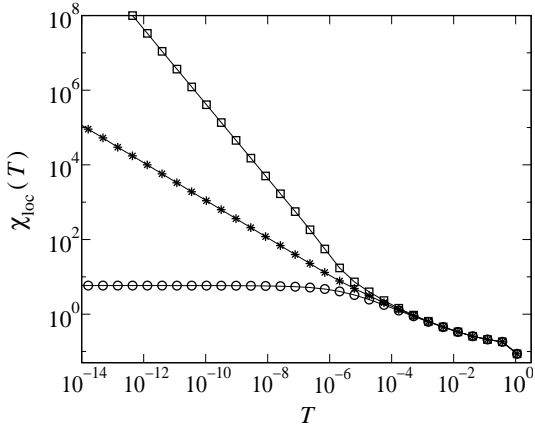}}
\caption{\label{NRG:chiloc}%
Static local susceptibilty $\cl(T)$ vs $T$ for $(r,s)=(0.2,0.5)$,
$\Gamma_0=0.1$, and $B_0=0.4902$ (circles), $0.5002\approx B_{0,c}$
(stars), and $0.5102$ (squares). The anomalous exponent in the
quantum-critical regime is found to be $\eta_{\chi}=1-s$,
independent of $r$. See text for further discussion.}
\end{figure}

We calculate the static local susceptibility via
\be
\cl(T)=-\left.\frac{\partial\mimp}{\partial\veps_f}\right|_{\veps_f=0}=
\lim_{\veps_f\to 0}-\frac{\mimp}{\veps_f}.
\ee
In the delocalized phase $B_0<B_{0,c}$,
$\mimp(T=0)$ vanishes
linearly with $\veps_f$ and thus $\cl(T)\approx \text{const.}$ for $T\ll T^*$.
In the localized phase $B_0>B_{0,c}$, $\mimp$ is nonzero as
$\veps_f \to 0$ with $\cl(T\ll T^*)\propto 1/T$. In the quantum-critical regime
$T^*\ll T \ll T_0$, $\cl(T)$ diverges as a power law with an anomalous exponent
$\eta_{\chi}$ defined in Eq.\ \eqref{chiloc}. For all $(r,s)$ pairs considered
(such that $\et>0$ and a critical fixed point exists), we find that
\be
\eta_{\chi}=1-s \, ,
\ee
independent of $r$. The behavior described above is clearly illustrated in
Fig.\ \ref{NRG:chiloc}, which shows three data sets for $(r,s)=(0.2,0.5)$:
one at the critical coupling and one close to it in either phase.
In this example, we extract $\eta_{\chi}=0.499(2)$.

\subsubsection{Hyperscaling}
\label{sec:NRGhyper}

As discussed in Sec.\ \ref{sec:RGhyper}, critical exponents for the present
model are expected to obey hyperscaling relations derived via a scaling ansatz
for the critical part of the free energy that assumes the critical fixed point
is interacting.\cite{insi} This expectation is borne out by the numerical
analysis: we find hyperscaling relations to be obeyed to within the estimated
error (typically less than 1\%) across the range of $(r,s)$ displaying critical
behavior. For example, for the case $(r,s)=(0.85,0.9)$, $1/\nu=0.082(1)$ and
$\eta_{\chi}=0.101(2)$. Thus, the values $\beta=0.601(2)$ and
$1/\delta=0.052(1)$ extracted from the data presented in Fig.\
\ref{NRG:deltabeta} obey Eqs.\ \eqref{betadelta} to within numerical
uncertainty.


\subsection{Spectral function}
\label{sec:NRGspectral}

We now turn to the single-particle spectral function $A(\w)$, calculated via
\be
A(\w)=\sum_{n,m}\left| \bra n | f^{\dagger} |m\ket \right|^2
\frac{e^{-\beta E_m}\!+\!e^{-\beta E_n}}{Z} \, \delta(\w-E_n+E_m) ,
\ee
where $|m\ket$ is a many-body eigenstate of NRG iteration $N$, and
$Z=\sum_n \exp(-\beta E_n)$ is the partition function; $A(\w)=A(-\w)$ for the
p-h symmetric parameters studied. The discrete delta-functions are Gaussian
broadened on a logarithmic scale: a standard NRG procedure discussed, e.g., in
Ref.\ \onlinecite{Bulla:07}. We set the broadening parameter $b$ such that
$A(\w)$ for the simplest resonant-level model (with $r=0$, $B_0=0$, and
$\veps_f=0$) is in optimal agreement with the exact result
$A(\w)=\pi^{-1}\Gamma_0/(\w^2+\Gamma_0^2)$.

\begin{figure}
\centerline{\includegraphics[clip,angle=0,width=5cm]{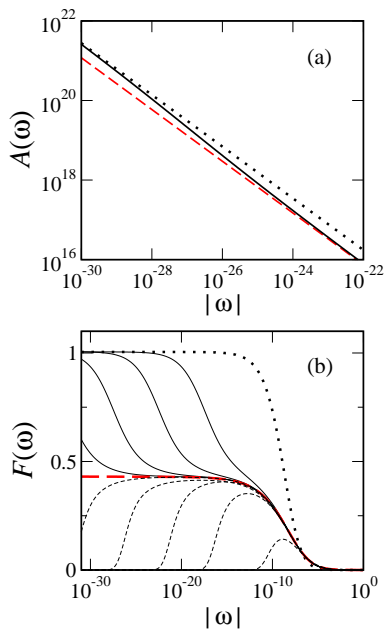}}
\caption{\label{spectral}%
(Color online) (a) Spectral function $A(\w)$ vs $|\w|$ for $r=0.65$, $s=0.8$,
$\Gamma_0=10^{-3}$, and three values of the dissipation strength: $B_0=0$
(dotted line), $B_0=B_{0,c}-10^{-5}$ (solid line), and
$B_0=B_{0,c}=0.03247113$ (thick dashed line).  At the quantum critical
point $B_0=B_{0,c}$, $A(\w)\propto|\w|^{-r}$, which behavior is also
followed for  $B_0$ close to $B_{0,c}$ and $|\w|\gg T_*$. In the
delocalized phase $B_0<B_{0,c}$, there is a crossover in $A(\w)$ to the
behavior Eq.\ \protect\eqref{A:deloc} for $|\w|\ll T_*$.  For the data
shown, $T^*\sim \mathcal{O}(10^{-26})$.  (b) The crossover behavior is
more readily seen in the modified spectral function
$\mathcal{F}(\w)=\pi\Gamma_0\sec^2(\frac{\pi}{2}r)|\w|^rA(\w)$,
which shows the ultimate low-$\w$ behavior $\mathcal{F}(\w=0)=1$
throughout the delocalized phase $B_0<B_{0,c}$, $0<\mathcal{F}(\w=0)<1$
for $B_0=B_{0,c}$, and $\mathcal{F}(\w=0)=0$ throughout the localized
phase $B_0>B_{0,c}$. In order of decreasing crossover scale,
delocalized-phase spectra are shown for $B_0=0$ (dotted line) and for
$B_{0,c}-B_0=$ $10^{-3}$, $10^{-4}$, $10^{-5}$, and $10^{-6}$ (solid lines);
localized-phase spectra (dashed lines) are shown for
$B_0=0.05$ and $B_0-B_{0,c}=$ $10^{-3}$, $10^{-4}$, $10^{-5}$, and
$10^{-6}$.  The critical spectrum is shown as a thick dashed line.}
\end{figure}

Figure \ref{spectral}(a) shows $A(\w)$ vs $|\w|$ on a logarithmic scale for
$r=0.65$, $s=0.8$, $\Gamma_0=10^{-3}$, and the dissipation strengths
$B_0\leq B_{0,c}$ specified in the figure caption. For the delocalized phase
$B_0<B_{0,c}$, we find that the dissipation does not alter the asymptotic
low-frequency behavior of $A(\w)$ found for $B_0=0$, i.e.,
\be
A(\w)=\frac{1}{\pi\Gamma_0} \, \cos^2 \!\! \left(\frac{\pi r}{2}\right)
|\w|^{-r} \ \ \ \text{for } |\w| \ll T^* .
\label{A:deloc}
\ee
For $B_0=0$ the spectrum is identical to that obtained for
the noninteracting ($U=0$) limit of the (spinful) pseudogap Anderson model at
p-h symmetry, where the result Eq.\ \eqref{A:deloc} holds for
$0<r<1$.\cite{GBI} Moreover, it is known\cite{Logan:00,Bulla:00} that
the form Eq.\ \eqref{A:deloc} persists throughout the Kondo-screened phase of
the pseudogap Anderson model with interactions present (i.e., for all $U<U_c$),
which in the p-h symmetric case is confined to $0<r<\half$.

In the vicinity of the QCP, $B_0\approx B_{0,c}$, we find
\be
A(\w)=\frac{\tilde{c}(r,s)}{\pi\Gamma_0} \,
\cos^2 \!\! \left(\frac{\pi r}{2}\right)|\w|^{-r} \ \ \
\text{for } T^*\ll |\w| \ll T_0 ,
\label{A:qcp}
\ee
where $\tilde{c}(r,s)\le 1$ and $T_0$ is a high-frequency
cutoff set by the bare hybridization strength $\Gamma_0$. This behavior
confirms Eqs.\ \eqref{tmatrix} and \eqref{etaTweak}.

In the localized phase, by contrast, $A(\w)$ vanishes as $\w \to 0$:
\be
A(\w)\propto |\w|^{a} \ \ \ \text{for } |\w| \ll T^* .
\label{A:loc}
\ee
The exponent $a$ is positive, and in general depends on both $r$ and $s$.

The crossover between these behaviors is more readily apparent in the modified
spectral function $\mathcal{F}(\w)=\pi\Gamma_0\sec^2(\pi r/2)|\w|^r A(\w)$.
Any low-frequency divergence of $A(\w)$ is canceled in $\mathcal{F}(\w)$, and
$\mathcal{F}(0)=1$ is pinned throughout the delocalized phase of the model.
As discussed in the context of the pseudogap Anderson
model,\cite{Glossop:00,Logan:00,Bulla:00}
this generalizes the well-known pinning $\pi\Gamma_0 A(0)=1$ of the spectral
function for the regular ($r=0$, fermionic) Anderson model. In the delocalized
phase, the scale $T^*$, playing the role of a renormalized tunneling amplitude,
is then manifest as the width of the pinned resonance at the Fermi level
$\w=0$, vanishing as $B_0\ra B_{0,c}^-$.

Figure \ref{spectral}(b) shows $\mathcal{F}(\w)$ vs $|\w|$ for $r=0.65$, $s=0.8$,
$\Gamma_0=10^{-3}$, and the $B_0$ values specified in the figure caption.
Throughout the delocalized phase ($0\le B_0 <B_{0,c}$), $\mathcal{F}(0)=1$
remains satisfied to within a few percent, as is typical for NRG. Close to the
QCP in either phase, $\mathcal{F}(\w)\approx \tilde{c}(r,s)$ down
to the scale $T^*$.

\begin{figure} \centerline{\includegraphics[clip,width=6cm,angle=270]{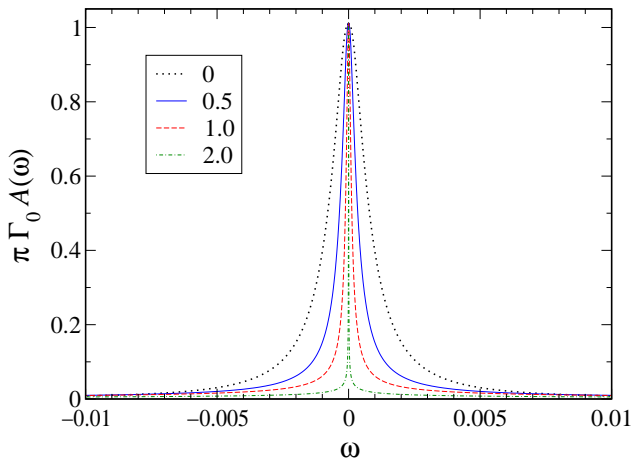}}
\caption{\label{spectral1}
(Color online)
$\pi\Gamma_0 A(\w)$ vs $\w$ for the case of a metallic fermionic
density of states ($r=0$) and Ohmic dissipation ($s=1$) for $\Gamma_0=0.001$.
Spectra are shown for increasing dissipation strength $B_0$ (see legend) in
the delocalized phase. The spectrum is a simple Lorentzian for $B_0=0$, and the
vanishing width as $B_0\ra B_{0,c}^-$ indicates a suppression of tunneling
between the local level and the conduction band.
}
\end{figure}

We close by considering the single-particle spectrum for the case of a metallic
fermionic density of states ($r=0$) and Ohmic dissipation ($s=1$). Here the
model describes charge fluctuations on a quantum dot or resonant tunneling
device close to a degeneracy point and subject to electromagnetic noise. The
essential physics---a Kosterlitz-Thouless-like QPT between delocalized and
localized states---has been investigated in a number of earlier studies
\cite{hur1,Borda:05,hur2,zarand,Glossop:07}, e.g., via a Bose-Fermi Kondo
model, and we will not repeat the discussion here.  We simply show, in  Fig.\ \ref{spectral1},
 the spectrum for $\Gamma_0=0.001$ and a range of dissipation
strengths; for $B_0=0$, $A(\w)$ is of Lorentzian form. The vanishing width of
the central resonance as $B_0\ra B_{0,c}^-$ indicates a suppression of
tunneling between dot and leads due to the noisy electromagnetic environment.

\section{Conclusions}

In this paper, we have analyzed the phase diagram and the quantum phase
transitions of a paradigmatic quantum impurity model with both fermionic
and bosonic baths, namely a dissipative resonant-level model.
For weak dissipation, the resonant tunneling of electrons is renormalized due
to the friction of the bosonic bath, but the ground state remains delocalized.
For strong dissipation, by contrast, the tunneling amplitude renormalizes to
zero in the low-energy limit leading to a localized ground state.
We have employed both analytical and numerical techniques, utilizing epsilon
expansions recently developed in the context of the pseudogap Anderson and
Kondo model, and an extension of Wilson's numerical renormalization-group
approach, generalized to treat both fermionic and bosonic baths.

The transition between delocalized and localized phases exists for a wide range
of exponents $r$ and $s$ characterizing the conduction-band and bosonic-bath
densities of states, respectively. Our epsilon expansions, formulated in the
original degrees of freedom, are in excellent agreement with numerics in the
vicinity of the expansion points. For the case of a metallic bath, inaccessible
to the analytical techniques used here, we have presented numerical results,
making contact with earlier bosonization studies of related models.

We finally mention a few applications. In the context of nanostructures,
a resonant-level model may describe the tunneling of electrons between a lead
and a small island or quantum dot.\cite{matveev,berman}
Taking into account electromagnetic noise of a fluctuating environment directly
leads to a model of type \eqref{H}, provided that the spin degree of freedom
of the electrons can be neglected (e.g., if electrons are spin-polarized due
to a large applied magnetic field).
Related situations, mainly corresponding to bath exponents $r=0$ and $s=1$,
have been discussed in the literature.\cite{hur1,hur2}
Apart from the common situation of ohmic noise ($s=1$), sub-ohmic dissipation
($s<1$) can occur, e.g., in RLC transmission lines which display a $\sqrt{\w}$
spectrum in the R-dominant limit.\cite{nazarov}
Further, a bath with $r=1$ may be realized using Dirac electrons of graphene
or quasiparticles of a $d$-wave superconductor.

\acknowledgments

We thank S.\ Florens and N.\ Tong for fruitful discussions on the present
paper and related subjects. This research was supported by the DFG through
the Center for Functional Nanostructures (Karlsruhe) and SFB 608 (K\"oln),
and by the NSF under Grant DMR-0312939.  C.H.C. acknowledges support
from the National Science Council (NSC) and the MOE ATU Program of
Taiwan, R.O.C.  We also acknowledge resources and
support provided by the Univ.\ of Florida High-Performance Computing Center.


\end{document}